\documentclass{article}
\usepackage{ismir,amsmath,cite,url}
\usepackage{amsfonts}
\usepackage{graphicx}
\usepackage{color}
\usepackage{enumitem}
\usepackage[nolist,nohyperlinks]{acronym}

\begin{acronym}
    \acro{GAS}{Google Audio Set}
    \acro{MAP}{Mean Average Precision}
    \acro{FMA}{Free Music Archive}
    \acro{LSA}{Latent Semantic Analysis}
    \acro{HR}{Hit Rate}
\end{acronym}

\title{Audio based disambiguation of music genre tags }

\oneauthor
  {Romain Hennequin, Jimena Royo-Letelier, Manuel Moussallam} {Deezer R\&D, Paris\\research@deezer.com}
  
\begin{document}

\maketitle
\begin{abstract}
In this paper, we propose to infer music genre embeddings from audio datasets carrying semantic information about genres.
We show that such embeddings can be used for disambiguating genre tags (identification of different labels for the same genre, tag translation from a tag system to another, inference of hierarchical taxonomies on these genre tags).
These embeddings are built by training a deep convolutional neural network genre classifier with large audio datasets annotated with a flat tag system.
We show empirically that they makes it possible to retrieve the original taxonomy of a tag system, spot duplicates tags and translate tags from a tag system to another.

%The abstract should be placed at the top left column and should contain about 150-200 words.
\end{abstract}
%

%\vspace{-1mm}
\section{Introduction}\label{sec:introduction}

%In this paper \emph{genre representation} can refer to a flat tag set of genre, a taxonomy which is a hierarchical structure of a set of concept, an ontology which is a general structure of a set of concept with arbitrary links between concepts, or a vector embedding where concept relations are encoded through proximity in the latent space.

% TODO  say that while vaguely define, genre remains a good proxy to access music.

Large genre annotated databases have been made available lately: the \ac{GAS} \cite{Gemmeke2017}, the MuMu dataset \cite{Oramas2017}, Discogs \cite{Bogdanov2017} or the \ac{FMA} dataset \cite{Defferrard2017} all contain hundreds of genre tags and hundreds of thousands multi-label genre track annotations.

Every dataset with genre annotations has its own genre representation: usually it is a tag set which is sometimes organized with a basic taxonomy (Discogs, MuMu, \ac{FMA}) or a basic ontology (\ac{GAS}).

However these representations usually suffer from ambiguity issues. First, tag definition may not be explicit: for the same tag name, definition may not be coherent from a dataset to another which prevents from doing correct translation from one tag set to another with a simple string matching.
Second, there may be duplicated tags \emph{i.e.} tag with different names but referring to the exact same genre such as \emph{Bossa Nova} and \emph{Bossanova} (without space) in Discogs.
Thirdly, there may be polysemy issues for some tags: it happens that a single tag refers to different concepts. In Discogs, the tag \emph{hardcore} may refer to \emph{hardcore punk} or to \emph{hardcore electronic music} which are quite different genres.
Finally, while a tag set may be structured in a taxonomy or an ontology, those have limitation for expressing all relations between tags: for instance the tag \emph{Blues Rock} in the MuMu taxonomy is a subgenre of \emph{Rock} and is not related to \emph{Blues}, which makes it as close to \emph{Electric Blues} as to \emph{Drum \& Bass} according to the taxonomy. Moreover taxonomy and ontology are generally designed with a particular purpose in mind \cite{Pachet2000}, possibly clarity for the customer for the MuMu taxonomy (which is the Amazon taxonomy), while it may be musicologic precision for DBpedia\footnote{wiki.dbpedia.org}, which may result in different meaning for tags and different relationship between them.

Building a genre representation from these tag systems in order to deal with these ambiguity issues can be done using a top-down approach, using an expert-level ontology such as the DBpedia ontology and trying to project the tag system into this ontology \cite{Diefenbach2016}. Mapping tags to an external expert ontology is not trivial, as a genre can have several different name and some tags may have several meanings: the tag \emph{funk} for instance may refer to a genre born in the 60s derived from soul and jazz, or, in Brazil, to Funk carioca which is a totally different style inspired by gangsta rap music. 
It also can be done using a bottom up approach, inferring relations between entities from data. The latter was mainly done using the genre tag distribution of a dataset with \ac{LSA} \cite{Sordo2008} or with a straight use of cooccurrences \cite{Schreiber2015, Schreiber2016} which all rely on the distributional hypothesis (similar tags are tags that cooccur a lot with same other tags). %, notably using total occurrences for each genre (that encodes popularity) and cooccurrences between tags (that encode similarity). 
However, it is sometimes not possible to rely only on tag distributions: the MuMu dataset has no overlap with the \ac{GAS}, which prevents from using tags cooccurrences to infer relationship between MuMu genre tags and \ac{GAS} ones.

So far, the literature has been mainly focusing on music genre classification on flat tag systems from audio \cite{Tzanetakis2002, Choi2016, Dieleman2011, Dieleman2014, Pons2016, Sanden2011}, text such as reviews \cite{Hu2005,Oramas2016} or lyrics \cite{Mayer2008, Choi2014},  album covers \cite{Libeks2011} or combinations of the previous modalities \cite{Neumayer2007,Schindler,Oramas2017},
while rarely addressing the actual semantic relationships that exist between genres.
%This is an issue since there exists no dataset with a fully satisfying genre ontology, and it is quite difficult to assess performance of such system in real world system not taking into account relations between tags: a classifier that confuses a lot \emph{Rock} and \emph{Classical Music} is certainly much worse than a classifier that confuses \emph{Electric Blues} and \emph{Modern Electric Blues}, even if both have similar accuracy metrics.
In \cite{Sturm2013}, %the authors pointed that \emph{``When a confusion table appears as a figure of merit, about $60\%$ of the time it is not accompanied by any kind of musicological reflection''}. So commenting musically the kind of representation learnt by the classification system is of great importance. 
the authors pointed out that focusing on classification metrics was not sufficient and suggested a deeper results analysis such as explanation of the confusion of the classifiers in term of musicological aspects. In this paper, we suggest going deeper in this direction and seeing how the confusion of the classifier is able to generate a structured genre representation: if the classifier is good enough, the confusion it makes should be able to encode the relation of proximity between genres.
Showing this property has two implications: it shows in a qualitative way that the classifier performs well and allows generation of a structured representation of a tag system using audio. 

In this paper, we thus aim to disambiguate genre tags and relations between them using audio as an alternative to the distributional hypothesis: we propose a method able to spot inconsistencies, help reducing them and relate tags between themselves, possibly across different non overlapping datasets with different tag systems. We enforce that the representation is based on audio only information and not on tag distribution using a monolabel learning scheme. While extracting a semantic representation from audio annotated with a flat tag representation was already sparsely addressed  (in \cite{Kolozali2013}, basic ontological relations between a few instruments are learnt back from isolated music instruments sounds and in \cite{Li2005} a simple music genre taxonomy is learnt with a few genre concepts), in this paper, we propose to learn representations at a large scale for tag systems with several hundreds of genre tags and with datasets of several hundreds of thousands of songs.
In Section \ref{sec:representation}, we explain how we compute genre tag embeddings using an audio-based genre classifier and use them to define an audio-based similarity between genre tags. In Section \ref{sec:validation}, we validate the learnt similarity by showing that it performs fairly on two artificial tasks (Discogs taxonomy learning and artificial deduplication). In Section \ref{sec:translation}, we show how we can use the learnt similarity to translate tags from a dataset tag system to another.
Finally, we draw conclusions in Section \ref{sec:conclusion}.

\vspace{-1mm}
\section{Learning a genre representation}
\label{sec:representation}
In this section, we explain how we build embeddings of genre tags using a genre classifier with audio input. We associate to each genre tag $t_i$ in the genre tag set $T = \{t_1,...,t_{N_c}\}$ an embedding vector ${\bf f}(t_i) = {\bf v}_{t_i} \in \mathbb{R}^n$, such that $d({\bf v}_{t_1},{\bf v}_{t_2})$ should correspond to an audio similarity between genre tag $t_1$ and genre tag $t_2$. %The embedding vector ${\bf v}_t$ is generated by audio-based genre classifier in different ways.
%The embeddings should have low distance when the genre are close according to the audio classifiers: the underlying assumption is that the embeddings should be related to the confusion between genres done by the classifier. %TODO: talk about that in the intro (or even abstract)

\vspace{-1mm}
\subsection{Datasets}

We use two large-scale genre annotated datasets for our experiments: The MuMu dataset \cite{Oramas2017}, and the genre provided by the Discogs website\footnote{https://discogs.com}.
We matched both datasets to Deezer track IDs using song metadata (album and artist names, and track titles).
We extracted a $30$s-long excerpt for each track (the position of the excerpt was sampled at random between the beginning and the end of the track). For tags with too few occurrences, we extracted several excerpts for balancing (as explained in Section \ref{sec:monolabel}). To avoid overlap between datasets we removed the $7260$ tracks that belong to both datasets (in order to not affect the translation experiment of Section \ref{sec:translation}).

While each dataset provides a simple genre taxonomy, we do not rely on it in the classification stage and consider the genre annotations as flat tag systems with no links between tags. The provided taxonomies are used afterwards for evaluation of the built genre representation.

\subsubsection{Discogs}

Discogs is referred as the ``\emph{largest open database containing explicit crowd-sourced genre annotations}'' in \cite{Bogdanov2017}.
It contains genre annotations at the album level for hundreds of thousands of albums. Genre tags in Discogs are organized in a two-level hierarchy: the first level, referred as \emph{genre}, includes generic genre categories (\emph{genre:Rock}\footnote{We prefix Discogs genre by \emph{"genre:"} to distinguish them from style}, \emph{genre:Jazz}, etc\ldots) and the second level, referred as \emph{style}, corresponds to subgenres (\emph{Psychedelic Rock}, \emph{Cool Jazz}, etc\ldots). 
It contains a total of more than $500$ genre/style tags. Only the $250$ most common tags were kept in our experiments ($235$ style tags and $15$ genre tags).

After cleaning, balancing (see Section \ref{sec:monolabel}) and matching, the Discogs dataset we used contained $418184$ tracks. %(some tracks could not be matched to Deezer tracks and then could not be used in the dataset)

\subsubsection{MuMu dataset}

The MuMu dataset \cite{Oramas2017} has genre annotation based on the Amazon $4$-level genre taxonomy. 
It contains genre annotations at the album level for $31471$ albums which contain a total of $147295$ tracks. 
It contains a total of $446$ different genres.
Only the top $211$ tags (those with less than $300$ annotated tracks are discarded) are kept.
After cleaning, balancing (see Section \ref{sec:monolabel}) and matching, the final MuMu dataset we used contained $122014$ tracks.

\subsubsection{Dataset split}

When training the system described in Section \ref{sec:classification_system}, we split the datasets into a training dataset ($70\%$), a validation dataset ($10\%$) used for early stopping, and a test dataset ($20\%$) used for building genre representations (Section \ref{sec:representation_learning}). The split was done at the artist level meaning two tracks by the same artist are in the same part of the split in order to avoid overfitting on variables such as album or artist as advised in \cite{Flexer2007AClassification,Pampalk2005ImprovementsClassificaton}.

\subsection{Monolabel learning}\label{sec:monolabel}

The annotations in a multilabel dataset carry information of popularity (through number of occurrences of a tag) and of similarity (through cooccurrences of tags).
This information was already used in several papers to build genre taxonomies from a flat tag system \cite{Schreiber2015,Schreiber2016,Sordo2008} or to build a target representation to improve classification results \cite{Oramas2017}.

The goal of the paper is to learn a genre representation only through audio and to avoid using non-audio information such as the one provided by the tag distribution.  As this distribution can be easily learnt as a side information in the last layer of a neural network, where bias can encode popularity (higher bias for more popular genre) while weights can encode similarity between genres (important value of dot product between weights corresponding to similar genres and vice versa), simply training a multilabel audio classifier based on a neural net will result in taking advantage of this information, and it may be difficult to assess at what point the actual audio information is relevant in building the representation from this classifier. 

In order to avoid influence of these non-audio information in the built genre representation, we propose to turn the multilabel classification problem into a monolabel one using the following learning scheme:
\begin{itemize}[leftmargin=*,parsep=0pt,itemsep=0pt,topsep=3pt]
\item To remove cooccurrences information, we transform the multilabel dataset into a monolabel one by sampling a tag among the multilabel tag annotation of every track.
\item To remove the popularity information, we balance equally all classes using a sampling probability inversely proportional to the global popularity of a tag (note, that it does not enforce perfect balancing).
\end{itemize}
For instance, if \emph{Rock} appears $1000$ times in the dataset and \emph{Punk} appears $100$ times, a song with (multi-)labels $\{$\emph{Rock}, \emph{Punk}$\}$ will get as monolabel \emph{Rock} with probability $1/11$ and \emph{Punk} with probability $10/11$. This ensures that rare genre tags have a high probability of being drawn, and that we keep the maximum of available information for rare tags while discarding somewhat redundant information for very common tags.

To enforce balancing, tags with too many occurrences are downsampled to keep a maximum of $2000$ occurrences per tag. Genre with not enough occurrences are upsampled to $2000$ occurrences by duplicating tracks (different $30$s excerpts are chosen for each track).

In order to avoid fitting independent variables, the sampling is done at the album level, which means that every track from the same album gets the same label. It also ensures that different excerpts of the same track have the same label.
Using this learning scheme, the confusion between genres should result only from similarities in audio.

\vspace{-1mm}
\subsection{Classification system}
\label{sec:classification_system}
We use a convolutional neural network with a recurrent layer on top of it as a monolabel classifier.
We feed it with Mel-spectrograms computed with $1024$ samples long Hann windows without overlap, with $96$ Mel filters. Audio is first downsampled to $22050$Hz and stereo channels are summed up. Mel-spectrogams were log compressed using the function $f(x) = \log(1 + Cx)$ where we chose $C=10000$. 
It results in $646\times96$ input matrices.

 The architecture of the neural network is quite similar to the one used in \cite{Choi2016} for automatic tagging, but with half as many filters in the convolutional layers (we noticed that it resulted in less overfitting) and a Gated Recurrent Unit \cite{Cho2014} on top of the conv layer (which improved overall classification accuracy). The gated linear unit was used for temporal pooling (only last temporal output is forwarded to the last layer which removes the time dimension) and was used in conjunction with dropout to reduce overfitting.
The architecture is summed up in \tabref{tab:CNN_architecture}.

The network was trained with a categorical cross-entropy loss with mini-batch stochastic gradient descent using Adadelta \cite{Zeiler2012}  %adaptive learning rate strategy 
and early stopping on the validation loss. The system was implemented with Keras \cite{Keras} using the Tensorflow \cite{Tensorflow} backend.

As the main goal of the paper is not to perform in terms of classification results, we did not try to optimize thoroughly the architecture and we just checked that our proposed system had similar classification results as in \cite{Oramas2017}.

\newcommand{\timesshort}{\!\!\times\!\!}%{\mkern-2mu\timesshort\mkern-2mu}}
\begin{table}
  \footnotesize
  \begin{center}
    \begin{tabular}{|l|l|l|}
        \hline
        Layer &         output shape  & N param.\\
        \hline
        Log-comp Mel-spec & $646\timesshort96\timesshort1$  & $0$\\
        Conv $3\timesshort3\timesshort64$ - MP $2\timesshort2$  & $323\timesshort48\timesshort64$ & $640$\\
        Conv $3\timesshort3\timesshort128$ - MP $3\timesshort4$    & $107\timesshort12\timesshort128$ &  $1280$\\
        Conv $3\timesshort3\timesshort256$ - MP $2\timesshort3$ & $53\timesshort4\timesshort256$ &  $2560$\\
        Conv $3\timesshort3\timesshort512$ - MP $3\timesshort4$  & $17\timesshort1\timesshort512$ &  $5120$\\
        GRU 512            & $512$   & $1574400$\\
        %Dropout $0.3$        & $512$   & $0$\\
        Dense Softmax      & $N_{c}$ & $512 \timesshort N_c$ \\
        \hline
    \end{tabular}
 \end{center}
\caption{Architecture of the Neural Network. MP stands for Max Pooling. %Activation of convolutionnal layer are rectified linear unit.
}
 \label{tab:CNN_architecture}
\end{table}

\vspace{-1mm}
\subsection{Genre embeddings from classification}
\label{sec:representation_learning}
There are several ways of extracting an embedding from a neural net based classification system. We describe the three kinds of genre embeddings we generated from the audio classifier in the following subsections. Whereas the first embedding only uses parameters of the classifiers, the other two make use of the test set. 

\vspace{-1mm}
\subsubsection{Last hidden layer weights}
The weights of the last hidden layer $\bf W$ are a $512\times N_c$ matrix. The $i$-th column of this matrix is then chosen as the embedding of genre tag $t_i$: 
\begin{equation}
    {\bf f}_w(t_i) = {\bf v}_{t_i} = {\bf W}_{:,i}.
    \label{eq:weight_embedding}
\end{equation}
This is a straightforward representation of a genre tag in the network: the output of the last hidden layer for a track annotated with some genre should be similar (in terms of dot product) to the weight vector of this genre. However, it necessitates retraining to incorporate new genre tags in the embedding.% Moreover, it is not stable in the sense that retraining the exact same network with the exact same dataset, but with different initialization can result in a totally different embedding function ${\bf f}_w$.

\subsubsection{Columns of output}
We can also build an embedding using the test dataset:
for every track $s$ in the test dataset, we denote $T_s$ the set of tags associated to $s$. 
We note the test dataset $S = \{s_1,s_2\ldots s_{N_s}\}$ where $s_k$ are the track excerpts.
The output of the network when fed with track excerpt $s_i$ is a vector ${\bf p}_k\in[0,1]^{N_c}$ (with $\sum_{j=1}^{N_c}  [{\bf p}_k]_j = 1$).
We note $\bf P$ the matrix in $\mathbb{R}^{N_s\times N_c}$ with ${\bf p}_k$ as $k$-th row.
The embedding of tag $t_i$ is then defined as the $i$-th column of matrix $\bf P$:
\begin{equation}
    {\bf f}_c(t_i) = {\bf v}_{t_i} = {\bf P}_{:,i}.
    \label{eq:columns_embedding}
\end{equation}
This embedding does not require annotation information about the tracks of the test set and the cosine similarity matrix between embeddings of all pairs of genre can be understood as a normalized confusion matrix and is the audio counterpart of the occurrence based representation defined in \eqref{eq:cooccurrence_embedding}. %It should be rather stable to retraining.
However it has very large dimension (that may be reduced using dimension reduction techniques such as \ac{LSA}) and it is quite difficult to add extra genres without retraining the whole system.
%- directly dependent on the dataset order which is a bad thing.

\subsubsection{Mean of output}

This third embedding type also uses the test dataset and takes advantage of the annotations. We note $S_{t_i} = \{s_{k_1}, s_{k_2}\ldots s_{k_{N_t}}\}$ the set of tracks annotated with genre tag $t_i$. We then associate to each $t_i$ the set of outputs of the classifier $\{{\bf p}_k | s_k \in S_{t_i}\}$. Ideally each genre tag $t_i$ would be represented by the distribution of all possible outputs for this genre. In practice, we compute statistics on these distributions. We then define the third type of genre tag embeddings as the mean of the output:
\begin{equation}
    {\bf f}_m(t_i) = {\bf v}_{t_i} = \frac{1}{|S_{t_i}|} \sum_{s_k \in S_{t_i}} {\bf p}_k.
    \label{eq:mean_embedding}
\end{equation}
As ${\bf p}_k$ is a categorical probability distribution, %$\forall k \in \{1,\ldots,N_c\}, {\bf p}_k\in[0,1]^{N_c}$ and $\sum_{j}  [{\bf p}_k]_j = 1$, 
%${\bf f}_m(t)\in[0,1]^{N_c}$ is also one ($\sum_{j}  [{\bf f}_m(t)]_j = 1$).
${\bf f}_m(t_i)$ is too.
%This third type of embedding should be stable (to retraining) and 
Embedding ${\bf f}_m$ makes it possible to incorporate new tags without retraining the whole system, by simply adding tracks annotated with the new genre tag in the dataset (the only constraints would be that the classifier was trained with similar genres): this is an important property of the embedding since it makes it much easier to incorporate new knowledge from another tag system.
%- should not be too much affected by sampling in the test dataset.

\subsubsection{Occurrence based representation}
In order to compare the audio-based representation we also define the following representation which is not based on audio but on tag distribution only. We note $\bf M\in\{0,1\}^{N_s\times N_c}$ the multilabel tag occurrence matrix with coefficient $M_{ij}=1$ iff track $s_i$ is annotated with tag $t_j$. The coocurrence embedding of tag $t_i$ is then defined as the $i$-th column of matrix $\bf M$:
\begin{equation}
    {\bf f}_{\text{dist}}(t_i) = {\bf M}_{:,i}.
    \label{eq:cooccurrence_embedding}
\end{equation}
This definition then shares similarity with the audio-based representation ${\bf f}_c$.

%\subsubsection{Characteristics of the proposed embeddings}

\subsubsection{Similarity measure}
\label{sec:cosine similarity}
To compare tags, we use the cosine similarity applied to the four types of genre tag embeddings defined in Equations \eqref{eq:weight_embedding}, \eqref{eq:columns_embedding}, \eqref{eq:mean_embedding} and \eqref{eq:cooccurrence_embedding}.% defined as:
%\begin{equation}
%    \text{cosim}({\bf u},{\bf v}) = \frac{{\bf u}^T{\bf v}}{\norm{\bf u}_2\norm{\bf v}_2}
%\end{equation}
%The similarity between two tags $t$ and $t'$ is then straighforwardly: $\text{cosim}({\bf v}_t,{\bf v}_{t'})$.

%In the following section, we show that the similarity based on the three proposed embeddings can be used to perform tags deduplication or translation and to retrieve hierarchical structure between tags. 
% TODO : add ?
%It is worth noting that, while we sometimes refer the similarity computed on ${\bf f}_{\text{dist}}$ as a cooccurrence similarity, two non cooccurring tags may have a quite high similarity if they often cooccur with the same set of tags.

\vspace{-1mm}
\section{Model Validation}
\label{sec:validation}
In this section, we validate that the audio-based similarities learnt in Section \ref{sec:representation} have a semantic meaning by showing that the original Discogs taxonomical relations can be inferred from the similarities and that they make it possible to spot duplicate tags in a dataset. In order to reproduce the results, we make available the embeddings, the similarity matrices we obtained for the different representation\footnote{\url{github.com/deezer/audio_based_disambiguation_of_music_genre_tags.git}} as well as dataset files (as lists of Deezer song IDs).

\vspace{-1mm}
\subsection{Taxonomy Learning}
\label{sec:taxonomy_learning}
In this section, we use similarity obtained from the genre embeddings described in Section \ref{sec:representation}, to infer hierarchical links between genres.
We trained the classification system with the Discogs dataset and the purpose of the experiment is to infer the genre/style links of the two-level Discogs taxonomy from audio.

The cosine similarity computed between genre tag embeddings provides a measure of similarity between genre tags. This can be used to rank for each style the similarity with each of the $15$ genres. The ground truth is the actual genre associated to the style in the Discogs taxonomy (note that some rare style are associated to $2$ music genres, such as \emph{hardcore} and \emph{noise} which are associated to both \emph{rock} and \emph{electronic}). 
We measure the quality of this ranking with classic ranking metrics: \ac{HR}@k which is the percentage of style for which the associated genre is in the top-$k$ according to the similiarity score.%  \ac{HR}@1 which is the percentage of style for which the similarity is maximum with the associated genre
(\ac{HR}@1 can be considered as a classification accuracy) and \ac{MAP} as defined in \cite{Zhu2004}.
\ac{MAP} takes into account the rank of the related genre in the similarity list.%: ranking as second a related genre is much better than ranking it as tenth, what \ac{MAP} takes into account while Hit \ac{HR}@1 does not.

Results are presented in \tabref{tab:taxonomy_learning_results}. As a reference, we report results for the occurrence based embedding ${\bf f}_{\text{dist}}$.  As style tags are always present together with their parent genre tag in the annotations, the performance of the occurrence based representation should be interpreted as an upper-bound for the results of the other representations, the errors being likely due to incoherences in the Discogs taxonomy (which is confirmed by the perfect \ac{HR}@2 score of ${\bf f}_{\text{dist}}$). Among the audio-based representations, ${\bf f}_c$ performs better than the two others. Despite being smaller than the occurrence based representation, we can see that the metrics for the audio representations are quite high, notably for ${\bf f}_c$ which has a near perfect \ac{HR}@2. This is noteworthy, since only audio information is used to infer the relations.%, while the similarity is based on audio information only. 

A qualitative analysis of the error shows that most of the ``errors'' (in the sense that the most similar genre to a style is not is related genre) actually make sense: for instance \emph{blues rock} which is a subgenre of \emph{genre:rock} in Discogs taxonomy has the greatest similarity (for ${\bf f}_c$) with \emph{genre:blues} which makes as much sense as the other (the same phenomena with hybrid subgenre appears with 
\emph{jazz-funk} and \emph{genre:funk / soul} instead of \emph{genre:jazz}, \emph{pop rock} and \emph{genre:rock} instead of \emph{genre:pop} and \emph{soul-jazz} and \emph{genre:funk / soul} instead of \emph{genre:jazz}). 
Other noteworthy examples are 
\emph{bossa nova} (subgenre of \emph{genre:jazz}) associated with \emph{genre:latin},
 \emph{musique concr\`ete} (subgenre of \emph{genre:electronic}) associated to \emph{genre:non-music} or \emph{rnb/swing} (subgenre of \emph{genre:hip hop}) associated to \emph{genre:funk / soul}. These qualitative results confirms that most of the ``errors'' are actually due to limitations of the original taxonomy and that \ac{HR}@2 may be the most revealing metric.
\begin{table}
 \footnotesize
 \begin{center}
   \begin{tabular}{|l|l|l|l|l|}
    \hline
    & ${\bf f}_w$ & ${\bf f}_c$ & ${\bf f}_m$ & ${\bf f}_{\text{dist}}$\\
    \hline
    \ac{HR}@1&$85.1$$\pm$$4.6$ & $89.4$$\pm$$3.9$ & $87.7$$\pm$$5.2$ & $96.2$$\pm$$2.5$\\
    \hline
    \ac{HR}@2 & $91.9$$\pm$$3.5$ & $98.3$$\pm$$1.7$ & $96.2$$\pm$$3.2$ & $100.0$$\pm$$0$\\
    \hline
%    \ac{MAP} & $0.906$\pm$0.029$ & $0.942$\pm$0.022$ & $0.883$\pm$0.030$ & $0.981$\pm$0.012$\\
    \ac{MAP} & $90.6$$\pm$$2.9$ & $94.2$$\pm$$2.2$ & $93.1$$\pm$$3.0$ & $98.1$$\pm$$1.2$\\
%   \hline  
%   \ac{MAP} & 0.906 & 0.942 & 0.882 & 0.981\\
    %\hline
    %AUC & 0.932 &  0.934 & 0.945\\
    \hline
  \end{tabular}
 \end{center}
 \caption{Average ranking metrics (in $\%$) for the Discogs taxonomy learning task with $95\%$ confidence intervals.}
 \label{tab:taxonomy_learning_results}
\end{table}
\begin{figure}
 \centerline{%\framebox{
 \includegraphics[width=\columnwidth]{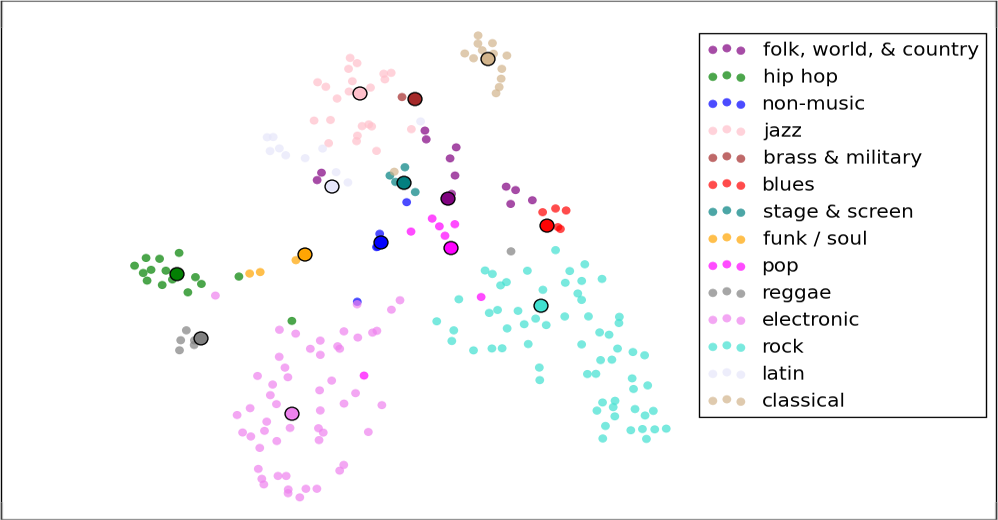}}%}
 \caption{$2$D t-SNE of ${\bf f}_w$ for the Discogs tags. Each \emph{style} is colored with the same color as its related \emph{genre}. %(style with several related genre were removed). 
 Main genres are depicted with bigger circle and black edges.}
 \label{fig:genre_tsne}
\end{figure}
In \figref{fig:genre_tsne}, we plot a $2$D t-distributed stochastic neighbor embedding (t-SNE) \cite{VanDerMaaten2008} of the learnt audio representation ${\bf f}_w$ in order to get visual insights about it: most music style tags are gathered in coherent clusters and are most of the time close to their related genre tag. A noteworthy exception is the style tags related to \emph{folk, world, \& country} that form several clusters, one of which being next to \emph{latin}, another one being next to \emph{blues} and another one next to \emph{pop}. This is pretty coherent since the tag \emph{folk, world, \& country} is supposed to gather several very different styles that may be closely related to other genres.

%We also tested multilabel classification on this task (without the monolabel sampling described in Section \ref{sec:monolabel}, with a sigmoid activation function on the output layer of the neural net and a binary crossentropy loss function) and it performed much better (\ac{HR}@1 of $95\%$ for the ${\bf f}_c$ embedding): this tends to confirms that a multilabel system is able to learn the underlying cooccurrence and popularity information of the tag system.
%It must be noted that for this task, using straightforwardly cooccurrences to infer the hierarchical style/genre links would lead to perfect results since style tags are always present together with their parent genre tag in the annotations, thus, it does not make sense to compare audio-based taxonomy inference with cooccurrence-based taxonomy for this task.
\vspace{-1mm}
\subsection{Tag deduplication}
\label{sec:deduplication}
In this section, we show how the audio-based similarities learnt in Section \ref{sec:representation} can be used to spot duplicates in a tag system.
To do that we rely on the ability of a classifier based on audio data to discriminate between two genre tags. If two genre tags cannot be discriminated, they probably have some strong relation (even if they have very dissimilar names). There may be several reasons for two tags having high confusion similarity: 
First, they may represent the exact same genre. Second, genre related audio characteristics may be very similar (the genre may be very similar with respect to audio). Thirdly, there may be differences of distribution in the datasets: datasets are usually an imperfect sample of the set of all music. Some genre may be biased toward a subgenre in a dataset while not in another one, which may result in strong differences in the meaning of some genres. Last, the classifier may not be able to distinguish them while there exists difference in some audio characteristics (that the classifier is not able to handle).%: this is the worst case.
\iffalse
\begin{itemize}[leftmargin=*,parsep=0pt,itemsep=0pt,topsep=3pt]
    \item They represent the exact same genre.
    \item Genre related audio characteristics are very similar (the genres are very similar with respect to audio).
    \item Difference of distribution in the database: databases are usually an imperfect sample of the set of all music. Some genres may be biased toward a subgenre in a dataset while not in another one. This may result in strong differences in the meaning of some genres.
    \item The classifier is not able to distinguish them while there exists difference in some audio characteristics (that the classifier is not able to handle).%: this is the worst case.
\end{itemize}
\fi

As it is very difficult to assess a ground truth for such a deduplication experiment, we propose the following artificial tag duplication: we use the Discogs genre dataset. We artificially duplicate every genre tag by creating two duplicate tags: for instance, \emph{Rock} is duplicated into \emph{Rock1} and \emph{Rock2}, which means that half of the tracks originally annotated with \emph{Rock} get the annotation \emph{Rock1} instead while the other half get the annotation \emph{Rock2}. To avoid learning the similarity through artist specific characteristics, we perform the split at the artist level, meaning that tracks of the same artist annotated with \emph{Rock} will get all the same subtag (either \emph{Rock1} or \emph{Rock2}).
Note that a subtag of group $1$ cannot cooccur  with 
a subtag of group $2$, which results in two separate tag systems (that we will refer as system $1$ and system $2$), with no overlap.
While all tags from system $1$ having a semantically equivalent counterpart in system $2$ is quite artificial, the total separation between the tag systems in term of cooccurrences is realistic. There is, for instance, no overlap between the \ac{GAS} and the MuMu dataset which means we can only rely on audio for linking them.

%The experiments aims at showing that an audio based genre representation is able to detect these duplicates.
In a similar way as in the experiment of Section \ref{sec:taxonomy_learning}, we use the similarity between genre tags embeddings as a \emph{duplication score}. The task is then for each genre tag, to retrieve its duplicated tag.
Once again, we present quantitative results in terms of \ac{HR}@k and \ac{MAP} in \tabref{tab:deduplication_results}. As opposed to the taxonomy learning task, it does not make sense to compare the audio based representations to the  occurrence-based representation since the sampling scheme we use avoid a tag of group $1$ cooccurring with a tag of group $2$ which means that the cosine similarity between any tag of group $1$ with any tag of group $2$ is $0$.
${\bf f}_w$ and ${\bf f}_c$ performs similarly, both performing significantly better than ${\bf f}_m$. Once again the score seems reasonably high for a representation based on audio information only.
% TODO: comment on low score of ${\bf f}_m$ => It is worth remindind that limited representation of distribution (mean only) while being the only proposed representation that makes it possible to add genre without retraining of the classifier.

It is interesting to look at the ``errors'' (when the most similar tag is not the actual duplicate) done by the system using ${\bf f}_c$. Some errors were actual duplicates in Discogs: \emph{bossa nova} was associated to \emph{bossanova} (without a space) which is clearly a duplicate issue in Discogs. Other example are \emph{style:reggae} and \emph{genre:reggae} (where a \emph{style} tag as the same name as its related \emph{genre} tag) or \emph{thug rap} and \emph{gangsta} (considered as the same genre in Wikipedia). This shows that the genre similarity computed from the embeddings is able to spot actual duplicates and that \ac{HR}@2 may be again the most revealing metric.
Some errors are matching between quite different concepts but with very similar audio, such as \emph{field recording}/\emph{musique concr\`ete},  %\emph{poetry}/\emph{genre:non-music}, 
\emph{poetry}/\emph{spoken word}, \emph{spoken word}/\emph{genre:non-music} and \emph{conscious}/\emph{genre:hip hop}.
Other errors are with very similar genres: \emph{bop}/\emph{hard bop}, \emph{honky tonk}/\emph{country blues}, \emph{space rock}/\emph{post rock}
A few errors are more difficult to explain such as
\emph{ragtime}/\emph{tango} which may have some audio similarities (the use of piano is quite common in both genres, and both are intended for dancing). % TODO : trouver un exemple qui ne marche pas trop.
These errors may come from the classification system we use or from a strong bias or annotation noise in the Discogs annotations.
\begin{table}
 \footnotesize
 \begin{center}
   \begin{tabular}{|l|l|l|l|}
    \hline
%    Embedding & ${\bf f}_w$ Eq. \eqref{eq:weight_embedding}& ${\bf f}_c$ Eq. \eqref{eq:columns_embedding}& ${\bf f}_m$ Eq. \eqref{eq:mean_embedding}\\
    & ${\bf f}_w$ & ${\bf f}_c$ & ${\bf f}_m$ \\
    \hline
    \ac{HR}@1 & $92.0\pm2.4$ & $92.8\pm2.3$ & $74.8\pm3.8$ \\
    \hline
    \ac{HR}@2 & $95.8$$\pm$$1.8$ & $97.0$$\pm$$1.5$ & $83.0$$\pm$$3.3$ \\
    %\hline
    %\ac{HR}@3 & $97.4\pm1.4$ & $98.2\pm1.2$ & $89.4\pm2.7$\\
    \hline
    \ac{MAP} & $98.1\pm0.6$ & $98.4\pm0.5$ &$93.3\pm1.1$\\ 
%    \hline
%   AUC & 0.999 &  0.941 & 0.998\\
    \hline
  \end{tabular}
 \end{center}
 \caption{Average ranking metrics (in \%) for the Discogs deduplication task with $95
 \%$ confidence intervals.}
 \label{tab:deduplication_results}
\end{table}

\vspace{-1mm}
\section{Tags translation}
\label{sec:translation}
In this section, we  perform another experiment that aims at  translating tags from MuMu dataset to Discogs dataset. For sack of clarity Discogs tags are prefixed with ``\emph{D:}'' and  MuMu tags with ``\emph{M:}''. When there are no or few overlaps between two datasets, we cannot rely on cooccurrences of tags to model relation between the tag systems. The only media we can rely on is then audio.

To train the classifier (see Section \ref{sec:classification_system}), we used the concatenation of the tags from the MuMu dataset and the Discogs dataset. Tags of each dataset were considered different even if they had the exact same name: \emph{e.g.}, there were a \emph{M:jazz} tag that was considered different from the \emph{D:jazz} tag. 
The experiment of translation is then very similar to the deduplication task presented in \ref{sec:deduplication}: the translation task consists of deduplicating the whole MuMu/Discogs tag set, focusing on pairs of duplicates for which the first element is a MuMu tag and the second element is a Discogs tag.

This allows to translate tags from one tag system to another, but also to spot possible genre definition differences between datasets: if two genre tags from two different datasets, with the exact same name can be discriminated with audio, this is probably because they do not carry the exact same meaning (provided we can move appart overfitting of the audio classifier used to build the representation).

We only consider here simple one-to-one tag mappings between MuMu and Discogs although it is restrictive since there may exist one-to-many mapping (e.g. between \emph{M:avant garde \& free jazz} and  \emph{D:avant-garde jazz}/\emph{D:free jazz}) or even more complex relationships.

\iffalse
% TODO : reintegrate ??
In order to get a very raw ground truth on a subset of the tags, we used a basic genre tag normalization as in \cite{Schreiber2015} and looked for exact string matches between MuMu tags and Discogs tags. This way, we matched $65$ tags (out of $211$ for MuMu and $250$ for Discogs).
Making the assumption that having (normalized) same name in both datasets means having the same meaning (which is not always true), this can be used as a ground truth for the MuMu/Discogs translation task.
\fi

% TODO put the definition of the cooccurrence matrix in definition of representations.
As the Discogs and MuMu datasets have some common tracks, we can compare the audio-based similarities with the cooccurrence-based one derived from ${\bf f}_{\text{dist}}$.

\begin{table}
\tiny
 \begin{center}
   \begin{tabular}{|l|l||l|l|}
   \hline
    \multicolumn{2}{|l||}{\scriptsize{Audio-based translation ${\bf f}_c$}} & \multicolumn{2}{|l|}{\scriptsize{Cooccurrence-based translation ${\bf f}_{\text{dist}}$}}\\
    \hline
    Mumu tag&              Discogs tag& Mumu tag&              Discogs tag\\
    \hline
         bebop & bop &irish folk  &            celtic \\
        movie scores &                 score&contemporary big band  &          big band \\
        indie \& lo-fi &                 lo-fi &latin music  &       genre:latin \\
        electric blues & modern elec. blues &rap \& hip-hop  &     genre:hip hop \\
        electronica &             leftfield &vocal blues  &           ragtime \\
        punk-pop &              pop punk &dance \& electronic  &  genre:electronic \\
        modern postbebop &            genre:jazz &today's country  &           country \\
        special interest &            avantgarde &electric blues  &       genre:blues \\
        singer-songwriters &             folk rock &children's music  &  genre:children's \\
        r\&b &             rnb/swing &comedy \& spoken word  &            comedy \\
    \hline
  \end{tabular}
 \end{center}
 \caption{Top $10$ most similar tags between MuMu and Discogs according to ${\bf f}_c$ (left columns) and  ${\bf f}_{\text{dist}}$ (right columns), removing string matched tags.
 }
 \label{tab:mumu_discogs_translation}
\end{table}

%Having seen the quantitative results on the string matched tag, 
There are two aspects that may be qualitatively assessed: why would two tags with different names be associated? and why would two tags with same name have a very low audio similarity.

In the two first columns of \tabref{tab:mumu_discogs_translation}, we present the $10$ Discogs tags that are most similar (according to ${\bf f}_c$) to MuMu tags while not having the same normalized name.
As can be seen, when the names are different, it can be due to the following reasons:
\begin{itemize}[leftmargin=*,parsep=0pt,itemsep=0pt,topsep=3pt]
    \item Two different names are used for the exact same concept: \emph{M:bebop}/\emph{D:bop}, \emph{M:punk-pop}/\emph{D:pop punk},
    \emph{M:movie scores}/\emph{D:score}.% (note that \emph{D:score} may also refer to videogame or television scores). %Other noteworthy examples in the top $20$ are \emph{swing jazz}/\emph{swing} and \emph{M:rap & hip-hop}/\emph{D:genre:hip hop}.
    \item Some genres were considered sufficiently similar to be grouped under the same tag name in one of the tag system while they were not in the other one \emph{e.g.}
    \emph{M:indie \& lo-fi}/\emph{D:lo-fi}, \emph{M:r\&b}\emph{D:rnb/swing}.
    %\item One is a deriving genre of the other: \emph{electronica}/\emph{downtempo}, \emph{big beat}/\emph{breakbeat}.
    \item One genre is a subgenre of the other: \emph{M:electric blues}/\emph{D:modern electric blues}, \emph{M:modern postbebop}/\emph{D:genre:jazz}
    %\item The genre are related: \emph{southern rap}/\emph{gangsta}
% Note that, according to Wikipedia\footnote{https://en.wikipedia.org/wiki/Southern_hip_hop}, \emph{Southern Rap} is a deriving genre of \emph{hardcore hip hop}, which was considered sufficiently similar to \emph{gangsta} to be grouped under the same tag name in MuMu.
\end{itemize}
The association between \emph{M:singer-songwriters} (a subgenre of \emph{M:rock}) with \emph{D:folk rock} (a subgenre of \emph{D:genre:rock}) seems to link quite similar concepts (which seems to be confirmed by the cooccurrence based similarity that is quite high).
\emph{M:electronica} and \emph{D:leftfield} seem to be quite broad electronic genres: the span of the former and the lack of precise definition of the latter while both seem not intended for dancing could explain the association.
The association \emph{M:special interest}/\emph{D:avantgarde} remains quite unclear, while the tags are quite vague.

In the two last columns of \tabref{tab:mumu_discogs_translation} are presented top $10$ most similar tags between MuMu and Discogs according to the cooccurrence based similarity (excluding string matched pairs with basic normalization as in \cite{Schreiber2015}). It can be seen that the top $10$ for cooccurrences and the top $10$ for audio similarity contains mostly different tags, with the exception of \emph{M:electric blues} which is not mapped to the same Discogs tag: this tends to show that cooccurrence similarity is complementary to the audio-based similarity, and when cooccurrence information is available (overlap between dataset), using both similarities should provide the best analysis.
% TODO : comment on examples of pairs for which audio find a match and cooccurrence not or vice versa.
This is confirmed with some MuMu/Discogs pairs such as \emph{M:bebop}/\emph{D:bop} and \emph{M:post hardcore}/\emph{D:post-hardcore} which seems to be perfect mapping and have very high audio similarity but very low (less than $0.1$) cooccurrence similarity. The low cooccurrence similarity may be explained by a lack of data for these tags.
%The relation \emph{ambient}/\emph{leftfield} is a bit more complicated to comment: there seems that both terms are catch-all tags for electronic music (as it is made explicit for the Discogs tag on the Discogs website \footnote{https://www.discogs.com/style/leftfield}).

On the other hand, it is also interesting to check tags with the exact same name in both datasets, but with quite low similarity score: the tags \emph{electronic}, \emph{instrumental} have very low similarity (according to both ${\bf f}_c$ and ${\bf f}_{\text{dist}}$) from one database to another.
\emph{D:electronic} refers to a generic term for describing all electronic music while this exact same concept seemed to be carried by \emph{M:dance \& electronic} in MuMu. \emph{M:electronic} is actually a subgenre of \emph{M:progressive} which is a subgenre of \emph{M:rock} and then has a very different meaning than the one in Discogs. \emph{instrumental} (which is not a genre by itself) is considered a subgenre of \emph{M:new age} and \emph{M:country} in the MuMu taxonomy and a subgenre of \emph{D:hip hop} in Discogs (while a large number of non-hip-hop songs seems to have the \emph{D:instrumental} tag). 

%\emph{hardcore} may refers to three very different genre, \emph{Hardcore punk}, \emph{Hardcore electronic music} 
%and \emph{Hardcore hip hop}.
%While in the MuMu dataset, \emph{hardcore} is clearly identified as a subgenre of \emph{alternative rock}, in Discogs, it is considered as a subgenre of both \emph{electronic} and \emph{rock} and thus is a reference to two different genres.
%A similar issue occurs with \emph{disco} which is a subgenre of \emph{funk / soul} in Discogs while it is a subgenre of \emph{dance \& electronic} in MuMu.

Thus, audio made it possible to spot significantly different genres that were represented by the exact same string. This highlights that the meaning of some genre may vary significantly from a database to another and that string matching can result in wrongly matched concepts.
%Beyond doing exact translations, the embeddings can be used by themselves as a genre representation: 

\vspace{-1mm}
\section{Conclusion}
\label{sec:conclusion}
In this paper we presented a way of learning genre embeddings from audio and showed that they are able to encode semantic similarities between genre tags: we showed that these embeddings were able to build genre taxonomies, to spot duplicates in a dataset or to translate genre from one tag set to another one.
In future works, we plan to explore extraction of structured representation of other tag types than genre (mood, instruments, country...) from audio and to exploit other datasets such as the \ac{FMA} dataset or \ac{GAS} to learn a more global representation.
We also plan to explore in more details how we can use several sources (audio, expert based ontology, string matching, cooccurrences) to build richer representation from flat tag systems.
%\cite{Prockup2015}

\vspace{-1mm}
\section{Acknowledgements}
\label{sec:acknowledgment}
The authors would like to thank Guillaume Salha for fruitful conversations and Matt Mould for proof-reading.

% For bibtex users:
\bibliography{ISMIRtemplate}

\begin{thebibliography}{10}

\bibitem{Bogdanov2017}
Dmitry Bogdanov and Xavier Serra.
\newblock {Quantifying Music Trends and Facts Using Editorial Metadata From the
  Discogs Database}.
\newblock In {\em International Society for Music Information Retrieval
  Conference}, pages 89--95, 2017.

\bibitem{Cho2014}
Kyunghyun Cho, Bart van Merrienboer, Caglar Gulcehre, Dzmitry Bahdanau, Fethi
  Bougares, Holger Schwenk, and Yoshua Bengio.
\newblock {Learning Phrase Representations using RNN Encoder-Decoder for
  Statistical Machine Translation}.
\newblock In {\em Empirical Methods in Natural Language Processing}, 2014.

\bibitem{Choi2014}
Kahyun Choi, Jin~Ha Lee, and J.~Stephen Downie.
\newblock {What is this song about anyway?: Automatic classification of subject
  using user interpretations and lyrics}.
\newblock In {\em ACM/IEEE Joint Conference on Digital Libraries}, pages
  453--454. IEEE, sep 2014.

\bibitem{Choi2016}
Keunwoo Choi, George Fazekas, and Mark Sandler.
\newblock {Automatic tagging using deep convolutional neural networks}.
\newblock In {\em International Society for Music Information Retrieval
  Conference}, pages 805--811, 2016.

\bibitem{Keras}
Fran\c cois Chollet.
\newblock Keras: Deep learning library for theano and tensorflow, 2015.

\bibitem{Defferrard2017}
Micha{\"{e}}l Defferrard, Kirell Benzi, Pierre Vandergheynst, and Xavier
  Bresson.
\newblock {FMA: A Dataset For Music Analysis}.
\newblock In {\em International Society for Music Information Retrieval
  Conference}, 2017.

\bibitem{Diefenbach2016}
Dennis Diefenbach, Pierre~Ren{\'{e}} Lh{\'{e}}risson, Fabrice Muhlenbach, and
  Pierre Maret.
\newblock {Computing the semantic relatedness of music genres using semantic
  web data}.
\newblock In {\em CEUR Workshop}, volume 1695, 2016.

\bibitem{Dieleman2011}
Sander Dieleman, Phil{\'{e}}mon Brakel, and Benjamin Schrauwen.
\newblock {Audio-based Music Classification with a Pretrained Convolutional
  Network}.
\newblock In {\em International Society for Music Information Retrieval
  Conference}, pages 669--674, 2011.

\bibitem{Dieleman2014}
Sander Dieleman and Benjamin Schrauwen.
\newblock {End-to-end learning for music audio}.
\newblock In {\em IEEE International Conference on Acoustics, Speech and Signal
  Processing}, pages 6964--6968, 2014.

\bibitem{Tensorflow}
Mart\'in~Abadi et~al.
\newblock Tensorflow: Large-scale machine learning on heterogeneous distributed
  systems, 2015.

\bibitem{Flexer2007AClassification}
Arthur Flexer.
\newblock {a Closer Look on Artist Filters for Musical Genre Classification}.
\newblock In {\em Ismir 07}, pages 341--344, 2007.

\bibitem{Gemmeke2017}
Jort~F Gemmeke, Daniel~P.W. Ellis, Dylan Freedman, Aren Jansen, Wade Lawrence,
  R.~Channing Moore, Manoj Plakal, and Marvin Ritter.
\newblock {Audio Set: An ontology and human-labeled dataset for audio events}.
\newblock In {\em IEEE International Conference on Acoustics, Speech and Signal
  Processing}, pages 776--780, 2017.

\bibitem{Hu2005}
Xiao Hu, JS~Downie, Kris West, and AF~Ehmann.
\newblock {Mining Music Reviews: Promising Preliminary Results.}
\newblock In {\em International Society for Music Information Retrieval
  Conference}, pages 536--539, 2005.

\bibitem{Kolozali2013}
\c{S}efki Kolozali, Mathieu Barthet, Gy{\"{o}}rgy Fazekas, and Mark Sandler.
\newblock {Automatic ontology generation for musical instruments based on audio
  analysis}.
\newblock {\em IEEE Transactions on Audio, Speech and Language Processing},
  21(10):1--14, 2013.

\bibitem{Li2005}
Tao Li and Midsunori Ogihara.
\newblock {Music genre classification with taxonomy}.
\newblock In {\em Acoustics, Speech, and Signal Processing}, pages 197--200,
  2005.

\bibitem{Libeks2011}
Janis Libeks and Douglas Turnbull.
\newblock {You can judge an artist by an album cover: Using images for music
  annotation}.
\newblock {\em IEEE Multimedia}, 18(4):30--37, apr 2011.

\bibitem{Mayer2008}
Rudolf Mayer, Robert Neumayer, and Andreas Rauber.
\newblock {Rhyme and Style Features for Musical Genre Classification By Song
  Lyrics}.
\newblock In {\em International Society for Music Information Retrieval
  Conference}, pages 337--342, 2008.

\bibitem{Neumayer2007}
Robert Neumayer and Andreas Rauber.
\newblock {Integration of Text and Audio Features for Genre Classification in
  Music Information Retrieval}.
\newblock In {\em Advances in Information Retrieval}, pages 724--727. 2007.

\bibitem{Oramas2016}
Sergio Oramas, Luis Espinosa-anke, Aonghus Lawlor, Xavier Serra, Horacio
  Saggion, Music~Technology Group, Universitat~Pompeu Fabra, and
  Universitat~Pompeu Fabra.
\newblock {Exploring Customer Reviews for Music Genre Classification and
  Evolutionary Studies}.
\newblock In {\em International Society for Music Information Retrieval
  Conference}, 2016.

\bibitem{Oramas2017}
Sergio Oramas, Oriol Nieto, Francesco Barbieri, and Xavier Serra.
\newblock {Multi-label Music Genre Classification from Audio, Text, and Images
  Using Deep Features}.
\newblock In {\em International Society for Music Information Retrieval
  Conference}, 2017.

\bibitem{Pachet2000}
Fran{\c{c}}ois Pachet and Daniel Cazaly.
\newblock {A Taxonomy of Musical Genres}.
\newblock In {\em Content-Based Multimedia Information Access Conference},
  pages 1238--1245, 2000.

\bibitem{Pampalk2005ImprovementsClassificaton}
Elias Pampalk, Arthur Flexer, and Gerhard Widmer.
\newblock {Improvements of Audio-Based Music Similarity and Genre
  Classification}.
\newblock In {\em Ismir}, volume~5, pages 634--637, 2005.

\bibitem{Pons2016}
Jordi Pons, Thomas Lidy, and Xavier Serra.
\newblock {Experimenting with musically motivated convolutional neural
  networks}.
\newblock In {\em International Workshop on Content-Based Multimedia Indexing},
  volume 2016-June, pages 1--6. IEEE, jun 2016.

\bibitem{Sanden2011}
Chris Sanden and John~Z. Zhang.
\newblock {Enhancing Multi-label Music Genre Classification Through Ensemble
  Techniques}.
\newblock In {\em ACM SIGIR Conference on Research and Development in
  Information Retrieval}, pages 705--714, New York, New York, USA, 2011. ACM
  Press.

\bibitem{Schindler}
Alexander Schindler and Andreas Rauber.
\newblock {An Audio-Visual Approach to Music Genre Classification through
  Affective Color Features}.
\newblock In {\em European Conference on Information Retrieval}, pages 61--67,
  2015.

\bibitem{Schreiber2015}
Hendrik Schreiber.
\newblock {Improving genre annotations for the million song dataset}.
\newblock In {\em 16th International Society for Music Information Retrieval
  Conference}, pages 241--247, 2015.

\bibitem{Schreiber2016}
Hendrik Schreiber.
\newblock {Genre Ontology Learning: Comparing Curated With Crowd-Sourced
  Ontologies}.
\newblock {\em 17th International Society for Music Information Retrieval
  Conference}, pages 400--406, 2016.

\bibitem{Sordo2008}
Mohamed Sordo, Oscar Celma, Mart{\'{i}}n Blech, and Enric Guaus.
\newblock {The Quest for Musical Genres: Do the Experts and the Wisdom of
  Crowds Agree?}
\newblock In {\em 9th International Conference on Music Information Retrieval},
  pages 255--260, 2008.

\bibitem{Sturm2013}
Bob~L. Sturm.
\newblock {Aalborg Universitet Classification Accuracy Is Not Enough
  Classification Accuracy Is Not Enough On the Analysis of Music Genre
  Recognition Systems}.
\newblock {\em Journal of Intelligent Information Systems}, 41(3):371--406,
  2013.

\bibitem{Tzanetakis2002}
George Tzanetakis and Perry Cook.
\newblock {Musical Genre Classification of Audio Signals}.
\newblock {\em IEEE Transactions On Speech And Audio Processing}, 10(5), 2002.

\bibitem{VanDerMaaten2008}
Laurens {Van Der Maaten} and Geoffrey Hinton.
\newblock {Visualizing high-dimensional data using t-sne}.
\newblock {\em Journal of Machine Learning Research}, 9:2579--2605, 2008.

\bibitem{Zeiler2012}
Matthew~D Zeiler.
\newblock {ADADELTA: An Adaptive Learning Rate Method}, 2012.

\bibitem{Zhu2004}
Mu~Zhu.
\newblock Recall, precision and average precision, 2004.

\end{thebibliography}

\end{document}